\documentclass[AMA,STIX1COL]{WileyNJD-v2}
\usepackage{dsfont}
\usepackage[utf8]{inputenc}
\usepackage{amsthm,amsmath}
\usepackage{bm}
\usepackage{moreverb}
\usepackage{booktabs}

\articletype{Discussion}

\received{26 April 2016}
\revised{6 June 2016}
\accepted{6 June 2016}

\begin{document}

\title{Broad versus narrow research questions in evidence synthesis: a parallel to (and plea for) estimands}

\author[1]{Antonio Remiro-Az\'ocar}

\author[2]{Anders Gorst-Rasmussen}

\authormark{REMIRO-AZ\'OCAR \& GORST-RASMUSSEN}

\address[1]{\orgdiv{Methods and Outreach}, \orgname{Novo Nordisk Pharma}, \orgaddress{\state{Madrid}, \country{Spain}}}

\address[2]{\orgdiv{Biostatistics HTA}, \orgname{Novo Nordisk A/S}, \orgaddress{\state{S\o borg}, \country{Denmark}}}

\corres{*Antonio Remiro-Az\'ocar, Methods and Outreach, Novo Nordisk Pharma, Madrid, Spain. \email{aazw@novonordisk.com}. Tel: (+34) 91 334 9800} 

\presentaddress{Antonio Remiro-Az\'ocar, Methods and Outreach, Novo Nordisk Pharma, Calle V\'ia de los Poblados 3, Madrid, 28033, Spain}

\abstract{
There has been a transition from broad to more specific research questions in the practice of network meta-analysis (NMA). Such convergence is also taking place in the context of individual registrational trials, following the recent introduction of the estimand framework, which is impacting the design, data collection strategy, analysis and interpretation of clinical trials. The language of estimands has much to offer to NMA, particularly given the ``narrow'' perspective of treatments and target populations taken in health technology assessment.}


\maketitle

\renewcommand{\thefootnote}{\alph{footnote}}

\newcommand{\specialcell}[2][c]{%
  \begin{tabular}[#1]{@{}l@{}}#2\end{tabular}}

\section{Introduction}\label{S1}

We appreciate the opportunity to discuss the review by Ades et al of the last twenty years of network meta-analysis (NMA).\cite{ades2024twenty} Their reflections on broad versus narrow research questions in evidence synthesis are particularly insightful. Indeed, there has been a transition from broad to more specific research questions in NMA.\cite{ades2024twenty} We draw parallels to the transition in the design and analysis of individual registrational trials, with the recent introduction of estimands.\cite{ICHEMA, kahan2024estimands} 

We focus on the ``narrow'' perspective taken in health technology assessment (HTA), where NMA has become a cornerstone methodology for informing policy decisions about reimbursement. Since policy decisions are made for a specific healthcare setting, the research questions guiding these decisions should be similarly narrow; targeting specific treatments and dosing regimens, aiming at clinically consistent populations, and based on outcomes measured at defined points in the disease pathway.

From this perspective, we agree with Ades et al that ``every effort should be made to reduce quantitative heterogeneity'' \cite{ades2024twenty} so that treatment effect estimates are maximally relevant to policy decisions. In our view, estimands are a relevant language for explicating and addressing key sources of quantitative heterogeneity in evidence synthesis and should be explicitly considered when planning NMAs.

\section{Estimands: a narrow view of research questions}\label{S2}

When designing registrational clinical trials, a central objective is to assess the benefit-risk profile of a given treatment, potentially enabling regulators to grant marketing authorization for such treatment. To support this, registrational trials are guided by principles from the International Council for Harmonisation (ICH) of technical requirements for pharmaceuticals for human use.\cite{mehrotra2016seeking} In 2019, the final version of the ICH E9 (R1) addendum was published.\cite{ICHEMA} This aims to align the scientific question posed by the study team, the trial design and the analytical approach, by introducing the notion of an estimand. Estimands describe the inferential target of a trial using five components: population, treatment, variable (endpoint), strategies for intercurrent events, and (population-level) summary effect measure.\cite{ICHEMA, kahan2024estimands}  

Estimands have been increasingly adopted by regulatory authorities and are required for submissions in Europe, United States, Canada and China.\cite{ICHimplementation} They are currently fundamental in the protocols and statistical analysis plans of registrational Phase III trials.\cite{homer2022early} Reporting guidelines like CONSORT and SPIRIT, which predate the estimand framework, have been updated to incorporate the ICH E9 (R1) recommendations.\cite{kahan2023reporting, kahan2023consensus} Consequently, the reporting of estimands is anticipated to become standard practice in scientific publications of registrational trials.

The introduction of estimands was driven by a perceived lack of clarity in the research questions addressed by registrational trials. Particularly concerning was the ambiguous handling of intercurrent events (IEs). These are post-baseline events, such as treatment switching or discontinuation, affecting the existence or interpretation of outcome measurements.\cite{keene2023estimands} Historically, the target of estimation in RCTs was not articulated beyond designations of ``intention-to-treat'' (ITT) or ``per-protocol'', with readers attempting to deduce the inferential target based on the statistical methods.\cite{keene2023estimands} This often led to misunderstandings about the application of treatment effect estimates to decision-making contexts. 

For instance, readers might have interpreted an ITT analysis as one that: (1) includes all randomized subjects according to their allocated treatment arm; and (2) reflects the use of treatment in routine clinical practice. However, analytical methods labeled as ITT might have: (1) excluded subjects who do not experience IEs; and (2) estimated effects in hypothetical scenarios where subjects do not experience IEs.\cite{clark2022estimands,mitroiu2020narrative, mitroiu2022estimation} Notable examples, when outcomes are set to missing following the IE, are mixed models for repeated measures and last observation carried forward. The former implicitly imputes the missing outcomes using data from subjects who do not experience the IE, while the latter does so explicitly, using the last available outcome prior to the IE. 

In addition to addressing a clinical question that is either ambiguous or misaligned with the original intent, such ad hoc imputation techniques make untestable assumptions. In the pre-estimands era, sensitivity analyses would be employed to gauge the range of treatment effect estimates, but often without explicating how these related to the primary analysis and clinical question. In this sense, the pre-estimands era resembled the ``broad'' situation described by Ades et al, where substantial and unspecific heterogeneity only allows for conclusions to be drawn about a range of effects.\cite{ades2024twenty} 

In contrast, estimands aim to reduce the heterogeneity stemming from ambiguous questions. They align with the ``narrow’’ view of evidence synthesis described by Ades et al,\cite{ades2024twenty} where minimizing heterogeneity optimizes the relevance of treatment effect estimates to the research question. The use of estimands encourages trialists to think holistically, aligning the objectives, design, data collection and analysis of their trial. Similarly, embracing a narrow view in evidence synthesis involves a comprehensive process that includes refined and standardized reporting, as well as the use of innovative estimation methods such as multi-level network meta-regression.

\section{Evidence synthesis should account for estimands}\label{S3}

Ades et al conclude that ``Every effort should be made to reduce quantitative heterogeneity, by carefully specifying the target population, by modelling and adjusting for study-related and reporting biases, by appropriately synthesizing outcomes reported in different ways and times''.\cite{ades2024twenty} This call to action is sufficiently inclusive to leave room for estimands, whose components are key sources of quantitative heterogeneity and are increasingly important in trial design and reporting.\cite{kahan2024estimands, kahan2023reporting,kahan2023consensus, keene2023estimands} The ICH E9 (R1) addendum, while lacking explicit guidelines for estimands in NMA, cautions against a ``na\"ive comparison between data sources, or integration of data from multiple trials without consideration and specification of the estimand that is addressed in each''.\cite{ICHEMA}

Within the evidence synthesis community, the uptake of estimands has been modest. This is not for lacking interest in the precise specification of research questions. Variations of the PICO (Population, Intervention, Comparator, Outcome) framework\cite{schardt2007utilization} are often used to define the scope of evidence syntheses and are recommended by the PRISMA statement and its NMA extension.\cite{moher2010preferred,hutton2016prisma} PICO allows for both broad and narrow research questions, tailored to the level of heterogeneity that the context can accommodate. Within systematic literature reviews, broad questions set the boundary for trial eligibility and inform the data extraction process. Conversely, payers and HTA bodies generally pose narrow research questions, mirroring the specific nature of policy decisions. In this setting, mixing ``apples and oranges’’ is discouraged.

Like estimands, PICOs can be specified at the trial level. Such PICOs are more precise than those in evidence synthesis; investigators directly shape the research question by controlling their trial's design, data collection, and analytical methods. Conversely, meta-analysts often depend on secondary data. This restricts the granularity of the research question, particularly in retrospective meta-analyses, where the decision to analyze data is made after some or all the trials have been conducted.

At the trial level, estimands enable setting more precise research questions than PICOs by incorporating two additional attributes: (1) IE strategies; and (2) the summary effect measure.\footnote{As opposed to the variable/endpoint in the estimand framework, which is the clinical outcome measured at the subject level, this is a summary measure used to compare outcomes between treatments at the population level.} PICOs can be thought of as sets of compatible estimands, with each PICO encompassing multiple estimand definitions. Historically, IE strategies and the summary effect measure might have been secondary in the formulation of research questions in evidence synthesis. With the advent of ICH E9 (R1), they should be regarded as fundamental components of any well-defined research question.

Regarding IE strategies, an informed discussion is required about which estimands are present in the evidence base and which are pertinent to stakeholders. The ``treatment policy'' strategy, where IEs are considered irrelevant to the definition of the treatment effect, might be tempting. This aligns with ITT, is commonly reported and can be estimated with minimal assumptions. However, defaulting to this strategy is inadvisable where the key priority is relevance to the decision-making context. 

Consider certain oncology RCTs, where subjects assigned to the control switch to the experimental arm (``rescue therapy'') following disease progression. Here, the treatment policy strategy has limited relevance for decision-making in healthcare systems where rescue therapy is unavailable. In such systems, patients under the current standard of care cannot initiate the experimental intervention because this is not yet available in the market. A hypothetical strategy for the treatment switching IE – had rescue therapy been unavailable post-progression – would be more relevant for decision-making. 

A critique of hypothetical estimands is that their estimation hinges on untestable modeling assumptions, potentially undermining their internal validity. The trade-off between internal and external validity requires a nuanced evaluation, yet this issue remains contentious. For example, the Institute for Quality and Efficiency in Healthcare in Germany is opposed to hypothetical strategies, arguing that these are not aligned with real-world clinical practice, where IEs do occur.\cite{morga2023intention} We argue against the automatic assertion of primacy of internal validity,\cite{cook2002experimental} particularly in HTA. A well-constructed hypothetical estimand may have greater external validity and advanced sensitivity analysis methods, e.g.,~tipping point analyses, can assess internal validity concerns.\cite{ratitch2013missing}

Another key attribute of estimands is the summary effect measure. Patients and physicians in clinical practice care more about conditional measures, which come closer to an ``individualized'' interpretation, assuming that relevant variables are conditioned on in the estimand definition. For population-level decisions about reimbursement, marginal measures are typically of greater interest.\cite{remiro2022target} Ades et al state that: (1) NMA relies on exchangeability; (2) exchangeability is not vulnerable to imbalances in ``purely prognostic'' factors; and (3) due to trial randomization, NMA treatment coefficients can be interpreted as causal effect estimates.\cite{ades2024twenty} These claims depend on the summary effect measure. 

For so-called non-collapsible summary measures, marginal treatment effects are affected by the distribution of purely prognostic variables, even in the absence of treatment effect heterogeneity across subjects.\cite{remiro2024transportability} Consequently, model-based covariate adjustment may be warranted to account for cross-study imbalances in purely prognostic variables.\cite{remiro2024transportability}  Moreover, the standard a posteriori checks used to detect heterogeneity, e.g.,~subgroup analyses and interaction tests,\cite{ades2024twenty} even if adequately powered, may be insufficient to discard variables for adjustment or to verify the exchangeability of marginal effects.

Where the target estimand is marginal, even if each of the included RCTs is protected from bias, the NMA may still be threatened by cross-study imbalances in purely prognostic factors. These would compromise the causal interpretation of treatment effect estimates in specific populations. The exchangeability assumption is tied to the summary effect measure.\cite{remiro2024transportability} When targeting marginal treatment effects, difference measures such as mean or risk differences reduce dependence on model-based covariate adjustment and are less likely to compromise exchangeability.\cite{remiro2024transportability, webster2023choice} Conversely, despite being routinely pooled in evidence syntheses, non-collapsible measures (e.g.,~log odds ratios and log hazard ratios) are arguably unappealing for exchangeability.\cite{xiao2022odds, didelez2022logic}

Finally, a hurdle when integrating estimands into evidence synthesis is the current lack of standardization.\cite{remiro2022some} As highlighted by Russek-Cohen, ``individual studies were not planned with similar estimands nor they were necessarily planned in anticipation of a meta-analysis.''\cite{russek2022discussion} Indeed, there may be discrepancies in the estimands reported across different studies, but the analytical methodology can provide a promising resolution. Ades et al outline how quantitative heterogeneity can be reduced for ``outcomes reported in different ways'' through multivariate normal random effects meta-analysis (MVNMA).\cite{ades2024twenty} When trials report different sets of estimands, e.g.,~if a relevant hypothetical estimand is reported in some trials but not in others, MVNMA could be used to account for the totality of evidence.   

\section{The price of being precise}\label{S4}

There is a paradox that is inherent to the pursuit of clarity, both for estimands and from the ``narrow'' perspective of evidence synthesis. The more we insist on precision about the context to which our inferences should apply, the greater our reliance on sophisticated terminology and methodology that may be counterproductive for a goal of clarity. In the context of evidence synthesis, a more laissez-faire approach to heterogeneity (e.g.,~risk-of-bias tools), where potential biases are documented and evidence downgraded accordingly, is convenient and may appear more transparent to decision-makers.

There is also a feasibility dimension that needs to be balanced with the scientific considerations. In the context of the future Joint Clinical Assessment,\cite{julian2022can} an essential part of the new European Union HTA regulation, a wide variability in policy questions (PICOs) between states is expected. If states were to ask very specific research questions, e.g.,~by requesting specific IE strategies or summary effect measures as part of the scope, this could threaten the consolidation of scopes across states.

However, concerns regarding complexity or feasibility should not detract from the value of estimands in evidence synthesis. Estimands can help explicate, in a systematic way, the elusive heterogeneity that arises when asking ambiguous questions. Moreover, it is a language that will be difficult to avoid given its presence in trial publications informing future evidence syntheses. Estimands can facilitate the identification, and even the quantitative resolution, of potential misalignment and biases arising from the synthesis of trials that address different research questions. Thus, they will play an important role in the efforts called for by Ades et al to minimize quantitative heterogeneity in evidence synthesis.\cite{ades2024twenty}

\section*{Acknowledgments}

Not applicable. 

\subsection*{Financial disclosure}

Funding agreements ensure the authors’ independence in writing and publishing the article.

\subsection*{Conflict of interest}

The authors are employed by Novo Nordisk and declare no conflicts of interest, as this research is purely methodological. The views expressed in this article do not necessarily represent those of Novo Nordisk. 

\subsection*{Data Availability Statement}

Data sharing is not applicable to this article as no datasets were generated or analyzed.

\subsection*{Highlights}

\paragraph{What is already known}

\begin{itemize}
\item There has been a shift from broad to more ``narrow'' research questions in the practice of network meta-analysis. 
\item Health technology assessments take a ``narrow'' perspective in order to inform policy decisions for specific healthcare settings, focusing on specific treatment regimens, patient populations and outcomes. 
\item To ensure that treatment effect estimates are maximally relevant to policy decisions, every effort should be made to reduce quantitative heterogeneity.
\end{itemize}

\paragraph{What is new}

\begin{itemize}
\item The convergence from broad to more specific research questions is also occurring in the context of individual registrational trials.
\item The recent introduction of the estimand framework is influencing the design, data collection strategy, analysis, and interpretation of clinical trials.
\item Estimands are becoming increasingly prevalent in publications for registrational trials, and will play a crucial role in informing future evidence synthesis work.
\item Estimands can help identify and quantitatively resolve potential misalignments that arise when synthesizing trials addressing different research questions.
\end{itemize}

\paragraph{Potential impact for RSM readers}

\begin{itemize}
\item The language of estimands offers significant benefits to network meta-analysis, particularly given the “narrow” perspective of treatments and target populations in health technology assessments. 
\item Estimands can clarify heterogeneity from ambiguous questions and help mitigate key sources of quantitative heterogeneity. They should be explicitly considered in the planning of meta-analyses. 
\end{itemize}
\bibliography{wileyNJD-AMA}


\end{document}